\documentclass[sigconf]{acmart}

\acmConference[]{}{}{}

\usepackage[xcolor=orange]{changes}

\usepackage[english]{babel} 
\usepackage{xcolor}
\usepackage{color}
\usepackage{tcolorbox}
\usepackage{amssymb}
\usepackage{pifont}

\usepackage{float}
\usepackage{newfloat,caption}

\usepackage{algorithm}

\usepackage{algorithmicx}
\usepackage{algpseudocode}

\usepackage[inline]{enumitem}

\usepackage{makecell}
\usepackage{amssymb}
\usepackage{ltl}
\usepackage{mathtools}
\usepackage{numprint}

\definecolor{myBlue}{RGB}{0,133,255}
\definecolor{myOrange}{RGB}{255,115,0}
\definecolor{myGreen}{RGB}{0,115,0}

\DeclareMathAlphabet{\mathrmbf}{\encodingdefault}{\rmdefault}{bx}{n}
\DeclareMathAlphabet      {\mathbfit}{OML}{cmm}{b}{it}

\definechangesauthor[name={Claudio Menghi},color=orange]{CM}
\definechangesauthor[name={Shiva Nejati},color=red]{SN}
\definechangesauthor[name={Khouloud Gaaloul},color=blue]{KG}
\usepackage[colorinlistoftodos,prependcaption,textsize=tiny]{todonotes}

\newboolean{showcomments}
\setboolean{showcomments}{true}
\ifthenelse{\boolean{showcomments}}
{\newcommand{\nb}[2]{
  \fcolorbox{black}{yellow}{\bfseries\sffamily\scriptsize#1}
  {\sf\small$\blacktriangleright$\textit{#2}$\blacktriangleleft$}
 }
 
}
{\newcommand{\nb}[2]{}
 
}

\usepackage{multirow}
\DeclareFloatingEnvironment[fileext=frm,placement={!ht},name=Listing,within=section]{listing}

\newcommand\ourtool{ARIsTEO}

\newcommand\resq[1]{
\noindent 
\fcolorbox{green!40!black}{green!5}{\noindent 
 \parbox{0.98\columnwidth}{\noindent  #1}}\\
}

\definecolor{lightgray}{gray}{0.75}

\newcommand{\tickness}{1pt}

\newcommand{\totaltimerunningrqone}{\numprint{4315567}}
\newcommand{\totaltimerunningrqonedays}{\numprint{99}}

\newcommand{\luxspace}{LuxSpace}
\newcommand{\luxcasestudy}{\texttt{SatEx}}

 \setcopyright{none}
\renewcommand\footnotetextcopyrightpermission[1]{}

\begin{document}

\title[Testing Compute-Intensive Cyber-Physical Systems]{Approximation-Refinement Testing of\\ 
Compute-Intensive Cyber-Physical Models:\\ An Approach Based on System Identification}

\author{Claudio Menghi}
\affiliation{
  \institution{University of Luxembourg}
  \country{Luxembourg}
}
\email{claudio.menghi@uni.lu}

\author{Shiva Nejati}
\affiliation{
  \institution{University of Luxembourg}
  \country{Luxembourg}
}
\email{shiva.nejati@uni.lu}

\author{Lionel C. Briand}
\affiliation{
  \institution{University of Luxembourg}
  \country{Luxembourg}\\
    \institution{University of Ottawa}
  \country{Canada}
}
\email{lionel.briand@uni.lu}

\author{Yago Isasi Parache}
\affiliation{
  \institution{LuxSpace}
  \country{Luxembourg}
}
\email{isasi@luxspace.lu}

\balance

\renewcommand{\shortauthors}{Trovato and Tobin, et al.}

\begin{abstract}
Black-box testing  has been extensively applied to test models of Cyber-Physical systems (CPS) since these models are not often amenable to static and symbolic testing and verification. Black-box testing, however,  requires to execute the model under test for a large number of candidate test inputs. This poses a challenge for a large and 
practically-important category of  CPS models, known as \emph{compute-intensive} CPS (CI-CPS) models, where a single simulation  may take hours to complete.  
We propose a novel approach, namely \ourtool , to enable effective and efficient testing of CI-CPS models. Our approach embeds black-box testing into an iterative approximation-refinement loop. At the start, some sampled inputs and outputs of the CI-CPS model under test are used to generate a surrogate model that is faster to execute and can be subjected to black-box testing. Any failure-revealing test  identified for the surrogate model is  checked on the original model. If spurious,  the test results are used to refine the surrogate model to be tested again. Otherwise, the test reveals a valid failure. 
We evaluated \ourtool\ by comparing it with S-Taliro, an open-source and industry-strength tool for testing CPS models. 
Our results, obtained based on 
 five publicly-available CPS models, show that, on average,  \ourtool\ is able to find $24\%$ more requirements violations than S-Taliro and is $31\%$ faster than S-Taliro in finding those violations.
We further assessed  the effectiveness and efficiency  of \ourtool\ on   
 a large industrial case study from the satellite domain. In contrast to S-Taliro, \ourtool\ successfully tested two different versions of this model and could  identify three requirements violations, requiring four hours, on average, for each violation.    
\end{abstract}

\maketitle

\section{Introduction}
\label{sec:intro}

A common practice in the development of Cyber-Physical Systems (CPS) is to specify CPS behaviors using executable and dynamic models ~\cite{liebel2018model,chowdhury2018automatically,mathworks}. These models support engineers in a number of activities, most notably in automated code generation and early testing and simulation  of CPS.  Recent technological advancements in the areas of robotics and autonomous systems have led to increasingly more complex CPS whose models are often characterized as \emph{compute-intensive}~\cite{chaturvedi2009modeling,arrieta2016search,Arrieta:2018:MBT:3205455.3205490,sagardui2017multiplex,gonzalez2018enabling}. 
Compute-Intensive CPS models (CI-CPS)  require a lot of computational power to execute~\cite{arrieta2019pareto} since they include complex computations such as dynamic, non-linear and non-algebraic mathematics, and further, they have to be executed for long durations in order to thoroughly exercise interactions between the CPS and its environment. For example, non-trivial simulations of an industrial model of a satellite system, capturing the satellite behavior for $24h$, takes on average around  $84$ minutes (\char`\~1.5 hours)~\cite{Luxspace}.\footnote{Machine \textsc{M1}: $12$-core Intel Core $i7$ $3.20$GHz $32$GB of RAM.}   The sheer amount of time required for just a single execution of  CI-CPS models significantly impedes testing and verification of these models since many testing and verification strategies require to execute the Model Under Test (MUT) for hundreds or thousands of test inputs.

Approaches to verification and testing of CPS models can be largely classified into \emph{exhaustive verification}, and \emph{white-box}  and \emph{black-box testing}. Exhaustive verification approaches often translate CPS models 
into the input language of model checkers or Satisfiability Modulo Theories (SMT) solvers. 
CPS models, however, may contain constructs that cannot be easily encoded into the SMT solver input languages.
For example, CPS models specified in the Simulink language~\cite{mathworks}  allow importing arbitrary C code via S-Function blocks or include other plugins (e.g., the Deep Learning Toolbox~\cite{deep}). In addition, CPS models typically capture continuous dynamic and hybrid
systems~\cite{alur:15}. Translating such modeling constructs into low-level logic-based languages is complex,  has to be handled on a case-by case basis and may lead to loss of precision which may or may not be
acceptable depending on the application domain. Furthermore, it is well-known that model checking such systems is in general undecidable~\cite{henzinger1998s,alur1995algorithmic,6064535}.
White-box testing uses the internal structure of the model under test  to specifically choose inputs that exercise different paths through the  model. 
Most white-box testing techniques aim to generate a set of test cases that satisfy some structural coverage criteria (e.g.,~\cite{10.1007/978-3-540-71493-4_27,dang2009coverage}). 
To achieve their intended coverage goals, they may rely either on SMT-solvers (e.g.,~\cite{kong2015dreach,gao2013dreal}) or on randomized search algorithms (e.g.,~\cite{Nejati:2019:TCS:3340872.3340874,matinnejad2013automated,dreossi2015efficient}). But irrespective of their underlying technique, coverage-guided testing approaches are not meant to demonstrate that CPS models satisfy their requirements.

More recently, falsification-based testing techniques have been proposed as a way to test CPS models with respect to their requirements~\cite{yaghoubi2017local,7963007,abbas2014functional,Nghiem:2010:MTF:1755952.1755983}. These techniques are black-box and aim to find test inputs violating system requirements. They are guided by  (quantitative) fitness functions 
that can estimate how far a candidate test is from violating some system requirement. Candidate tests are sampled from the search input space using randomized or meta-heuristic search strategies (e.g.,~\cite{8453180,matinnejad2013automated,Matinnejad:2016:ATS:2884781.2884797}). To compute fitness functions, the model under test is executed for each candidate test input. The fitness values then determine whether the goal of testing is achieved (i.e., a requirement violation is found) or further test candidates should be selected. In the latter case, the fitness values may guide selection of new test candidates.  Falsification-based testing has shown to be effective in revealing requirements violations in complex CPS models that cannot be
 handled by alternative verification methods. However, serious scalability issues arise when testing CI-CPS models since simulating such models for every candidate test may take such a large amount time to the extent that testing becomes impractical.

In this paper, in order to enable efficient and effective testing of CI-CPS models, we propose a technique that  combines falsification-based testing with an approximation-refinement loop. Our technique, shown in  Figure~\ref{fig:ourapproach},   is  referred to as AppRoxImation-based TEst generatiOn (\ourtool).   As shown in the figure, provided with a  CI-CPS model under test (MUT), we automatically create an approximation of the MUT that closely mimics its behavior  but is significantly cheaper to execute. We refer to the approximation model as  \emph{surrogate} model, and generate it using System Identification (SI)  (e.g.,~\cite{soderstrom1989system,bittanti2019model}) which is a methodology for building mathematical models of dynamic systems using measurements of the system's inputs and outputs~\cite{soderstrom1989system}. Specifically, we use some pairs of inputs and outputs from the MUT to build an initial surrogate model. We then apply falsification testing to the surrogate model instead of the  MUT until we find a test  revealing some requirement violation for the surrogate model. The identified failure, however, might be spurious. Hence, we 
check the test on the  MUT. If the test is spurious,  we use the output of the test to retrain, using SI, our surrogate model  into a new model that more closely mimics the behavior of the MUT,  and continue with testing the retrained surrogate model.  If the test is not spurious, we have found a requirement violation by running the MUT very few times. 

\ourtool\ is inspired, at a high-level, by the counter-example guided abstraction-refinement (CEGAR) loop~\cite{kurshan1994computer,clarke1994model,clarke2000counterexample} proposed to increase scalability of formal verification techniques. 
In CEGAR, boolean abstract models are generated and  refined based on counter-examples produced by model checking, while in \ourtool, numerical approximation of CPS models are learned and retrained using test inputs and outputs generated by model testing. 

Our contributions are as follows:

\begin{figure}
\includegraphics[width=\columnwidth]{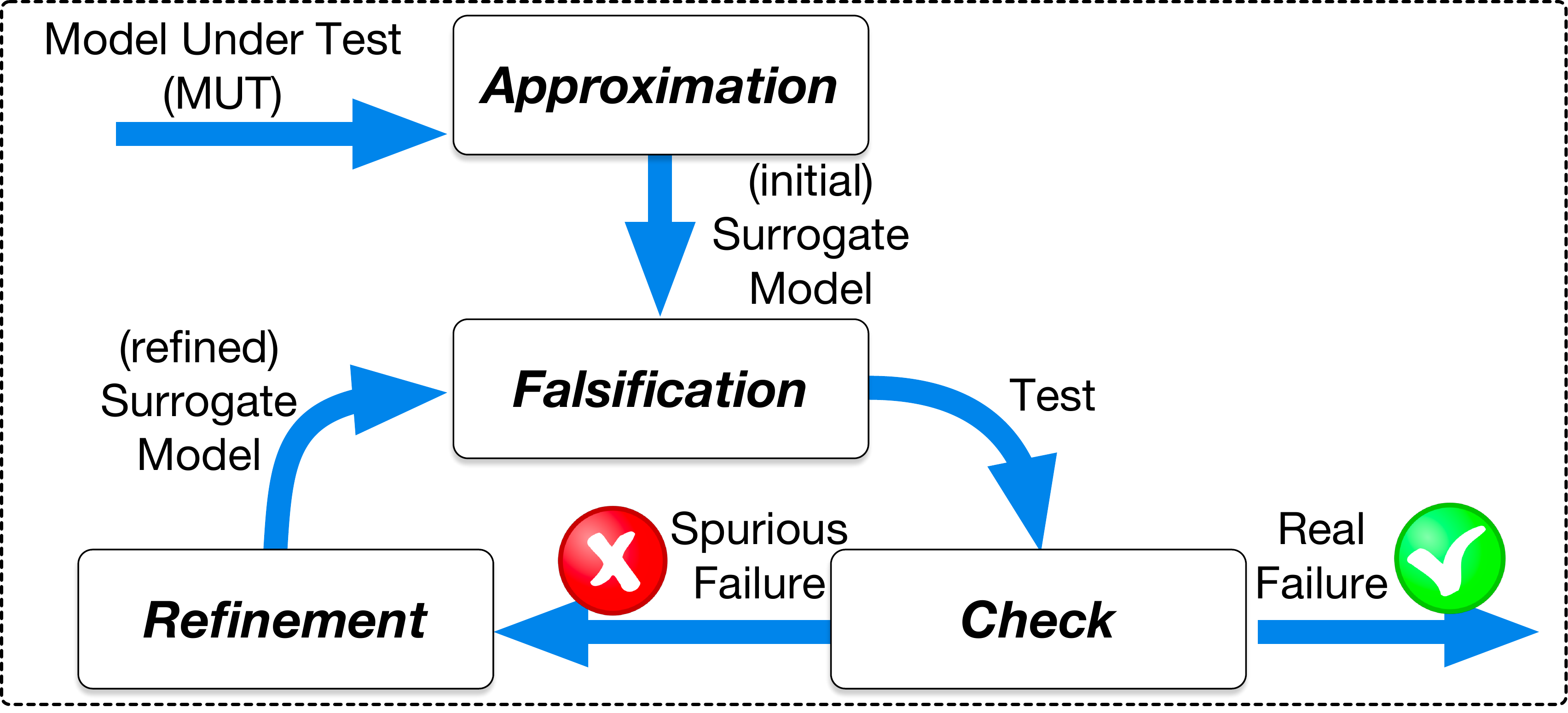}
\vspace{-0.7cm}
\caption{\ourtool : AppRoxImation-based TEst generatiOn.}
\label{fig:ourapproach}
\vspace{-0.3cm}
\end{figure}

\noindent $\bullet$ \emph{We 
developed \ourtool , an approximation-refinement testing technique, to identify  requirements violations for CI-CPS models.} \ourtool\  combines falsification-based testing with surrogate models built using System Identification (SI). We have implemented   \ourtool\  as a Matlab/Simulink standalone application, relying on the existing state-of-the-art  System Identification toolbox of Matlab as well as S-Taliro~\cite{staliro}, a state-of-the-art, open source falsification-based framework for Simulink models.

\noindent $\bullet$ \emph{We compared  \ourtool\ and S-Taliro to assess the effectiveness and efficiency of our proposed approximation-refinement testing loop.}  Our experiments, performed on five publicly-available Simulink models from the literature, show that, on average,   \ourtool\  finds $23.9\%$  more requirements violations than S-Taliro and finds the violations in $31.3\%$ less time than the time S-Taliro needs to find them.

\noindent $\bullet$ \emph{We evaluated usefulness and applicability of  ARIsTEO in revealing requirements violations in large and  industrial CI-CPS models from the satellite domain.} 
We analyzed three different requirements over  two different versions of a CI-CPS model provided by our industrial partner. \ourtool\ successfully detected violations in each of these versions and for all the requirements, requiring four hour, on average, to find each violation. In contrast, S-Taliro was not able to find any violation on neither of the model versions and after running for four hours.

\textbf{Structure.} 
Section~\ref{sec:running} presents our running example, formulates the problem and describes our assumptions.
Section~\ref{sec:ourtool} describes \ourtool , which is then evaluated in Section~\ref{sec:evaluation}.
Section~\ref{sec:related} presents the related work. Section~\ref{sec:conclusion} concludes the paper.

\section{CPS Models and Falsification-Based Testing}
\label{sec:running}

In this section, we describe how test inputs are generated for black-box testing of CPS models. We then introduce  the baseline falsification-based testing framework  we use in this paper to test CPS models against their requirements.

\textbf{Black-box testing of CPS models.} We consider CPS models under test (MUT) specified in Simulink  since  it is a prevalent language used in CPS development~\cite{liebel2018model,Dajsuren}.   
Our approach is  not tied to the Simulink language, and can be applied to  other executable languages requiring inputs and generating outputs that are signals over time (e.g., hybrid systems~\cite{grossman1993hybrid}). Such languages are common for CPS as engineers need to describe models capturing interactions of a system with its physical environment~\cite{alur:15}. We use \luxcasestudy, a model of a satellite,  as a running example. \luxcasestudy\ is a case study from our industrial partner, \luxspace~\cite{Luxspace}, in the satellite domain.

Let  \emph{time domain} $T=[0,b]$ be a non-singular bounded interval of $\mathbb{R}$.  A \emph{signal} is a function $f: T \rightarrow \mathbb{R}$. We indicate individual signals using lower case letters, and  \emph{sets} of signals using upper case letters. Let $\mathcal{M}$ be a MUT. We write $Y=\mathcal{M}(U)$ to indicate that the model $\mathcal{M}$ takes a set of signals $U=\{u_1, u_2 \ldots u_m\}$ as input and produces  a set of signals $Y=\{y_1, y_2 \ldots y_n\}$ as output. 
Each $u_i$ corresponds to one model input signal, and each $y_i$ corresponds to one model output signal.  We use the notation  $u_i(t)$ and $y_i(t)$ to, respectively, indicate the values of the input signal $u_i$ and the output signal $y_i$ at time $t$. For example, the \luxcasestudy\ model has four  input signals indicating the temperatures perceived by the Magnetometer, Gyro, Reaction wheel and Magnetorquer components, and one  output signal representing  the orientation (a.k.a attitude) of the satellite.

To execute a Simulink MUT $\mathcal{M}$, the simulation engine receives signal inputs defined over a time domain and computes signal outputs at successive time steps over the same time domain used for the inputs. A test input for $\mathcal{M}$ is, therefore,  a set of signal functions assigned to the input signals $\{u_1, u_2 \ldots u_m\}$ of $\mathcal{M}$.  To generate signal functions, we have to generate values over the time interval $T=[0,b]$. This, however, cannot be done in a purely random fashion, since input signals are expected to conform to some specific shape to ensure dynamic properties pertaining to their  semantic. For examples, input signals may be constant, piecewise constant, linear, piecewise linear, sinusoidal, etc. To address this issue, we parameterize each input signal $u_i$ by an interpolation function, a value range $R$ and a number $n$ of control points (with $n>2$).  To generate a signal function for $u_i$, we  then randomly select $n$ control points $u_i(t_1)$ to $u_i(t_{n})$  within $\mathbb{R}$  such that  $t_1=0$, $t_n=b$ and $t_2$ to $t_{n-1}$ are from $T$ such that $t_1 < t_2 < \ldots <  t_{n-1} < t_{n}$.
The values of  $t_2 < t_3 < \ldots < t_{n-1}$ can be either randomly chosen or they can be fixed with equal differences between each subsequent pairs, i.e,  $(t_{i+1}-t_i) = (t_{i}-t_{i-1}) $. The interpolation function is then used to connect  the $n$ control points  $u_i(t_1)$ to $u_i(t_n)$.
\ourtool\ currently supports several interpolation functions, such as 
piecewise constant,
linear  and piecewise cubic interpolation.
For each input $u_i$ of $\mathcal{M}$, we define a triple $\langle \mathit{int}_i, R_i, n_i \rangle$, where $\mathit{int}_i$ is an interpolation function, $R_i$ is the range of signal values and $n_i$ is the number of control points. We refer to the set of all such triples for all inputs $u_1$ to $u_m$ of $\mathcal{M}$ as  an \emph{input profile}  of $\mathcal{M}$ and denote it by $\texttt{IP}$. Provided with an input profile for a MUT $\mathcal{M}$, we can randomly generate test inputs for $\mathcal{M}$ as sets of signal functions for every input $u_1$ to $u_m$. For example,  the input profile for  \luxcasestudy\ provided by \luxspace\ is reported in Table~\ref{table:rq3}, where  $[-20,50]$, $[-15,50]$, $[-20,50]$, $[-20,50]$ are real value domains.

\textbf{Baseline falsification-based testing.} 
The goal 
 is to produce a test input $U$ that, when executed on the MUT $\mathcal{M}$, reveals a violation of some requirement of $\mathcal{M}$. Algorithm~\ref{algo:fal} represents a high-level overview of falsification-based testing. It is a black-box testing process and includes three main components: (1)~a test input generation component (\Call{Generate}{} in Algorithm~\ref{algo:fal}), (2)~a test objective determining whether, or not,  a requirement violation is identified  (\Call{TObj}{} in Algorithm~\ref{algo:fal}), and (3)~a search strategy to traverse the search input space and select candidate tests (\Call{Search}{} in Algorithm~\ref{algo:fal}).

\begin{table}[t]
\caption{Input Profile for the \luxcasestudy\ case study.
} 
\vspace{-0.3cm}
\label{table:rq3}
\scalebox{.8}{\begin{tabular}{c   c  c  c  c   }
\toprule
& \bf  Magnetometer & \bf  Gyro & \bf  Reaction wheel & \bf  Magnetorquer \\ 
\midrule
$\mathit{int(n)}$  & \texttt{pchip(16)} & \texttt{pchip(16)} & \texttt{pchip(16)} & \texttt{pchip(16)} \\
\midrule
$R$  & [-20,50] & [-15,50] & [-20,50] & [-20,50] \\
\bottomrule
\end{tabular}}
\vspace{-0.3cm}
\end{table}

We describe \Call{Generate}{}, \Call{Search}{} and \Call{TObj}{}. 
The input to the algorithm is a MUT $\mathcal{M}$ together with its input profile $\texttt{IP}$ and the maximum number $\texttt{MAX}$ of executions of MUT that can be performed within an allotted test budget time. Note that we choose the maximum number of executions as a loop terminating condition, but an equivalent terminating condition can be defined in term of maximum execution time.

\emph{Initial test Generation} (\Call{Generate}{}). It produces a (candidate) test input $U$ for $\mathcal{M}$ by randomly selecting control points within the 
ranges 
and applying the interpolation functions as specified in $\texttt{IP}$.

\emph{Iterative search} (\Call{Search}{}). It  selects a new (candidate) test input $U$ from the search input space of $\mathcal{M}$. It uses the input profile $\texttt{IP}$ to generate new test inputs. The existing candidate test input $U$ may or may not be used in the selection of the new test input. In particular, $\Call{Search}{\mathcal{M}, \texttt{IP},U}$ can be implemented using different randomized or meta-heuristic search algorithms~\cite{Nghiem:2010:MTF:1755952.1755983,matinnejad2015search,FSE2019Comp}. These algorithms can be purely \emph{explorative}  and generate the new test input randomly without considering the existing test input $U$ (e.g., Monte-Carlo search~\cite{Nghiem:2010:MTF:1755952.1755983}), or they may be purely \emph{exploitative} and generate the new test input by slightly modifying  $U$ (e.g., Hill Climbing~\cite{matinnejad2015search,FSE2019Comp}). Alternatively, the search algorithm may combine both explorative and exploitative heuristics (e.g., Hill Climbing with random restarts~\cite{luke2013essentials}).

\algrenewcommand\algorithmicindent{0.4em}
 \begin{algorithm}[t]
\caption{Baseline Falsification-based Testing.}
\label{algo:fal}
\begin{algorithmic}[1]
\Function{Falsification-Test}{$\mathcal{M}$,  $\texttt{IP}$, $\texttt{MAX}$}
\Repeat
\If{$U$ is null}
\State $U$ = \Call{Generate}{$\mathcal{M}$,  $\texttt{IP}$};  \Comment{Generate a candidate test input}
\Else
\State $U$ = \Call{Search}{$\mathcal{M}$,  $\texttt{IP}$, $U$};  \Comment{Generate next candidate test input}
\EndIf
	\State $Y = \mathcal{M}(U)$;  \Comment{Execute $\mathcal{M}$ for $U$}
    	\If{ $\Call{TObj}{U,Y} \leq 0$}  \Comment{Check if $U$ reveals a violation}
	\State  \textbf{return} $U$; 
	 \EndIf 
\Until{the number of executions of $\mathcal{M}$ reaches $\texttt{MAX}$}
\State	\Return NULL; \label{ln:noinputfals}
\EndFunction
\end{algorithmic}
\end{algorithm}

\emph{Test objective} (\Call{TObj}{}). It maps  every test input $U$ and its corresponding output  $Y$, i.e., $Y = \mathcal{M}(U)$, into a test objective value $\Call{TObj}{U, Y}$ in the set $\mathbb{R}$ of real numbers.  Note that computing test objective values requires simulating $\mathcal{M}$ for each candidate test input.  We assume for each requirement of $\mathcal{M}$, we have a test objective  $\Call{TObj}{}$  that satisfies the following conditions:  
\begin{enumerate}[parsep=0pt,itemindent=0cm,leftmargin=3\parindent]
\item[$\Call{TObj}{}1$] If  $\Call{TObj}{U, \mathcal{M}(U)}<0$, the requirement is violated;
\item[$\Call{TObj}{}2$] If  $\Call{TObj}{U, \mathcal{M}(U)} \geq 0$,   the requirement is satisfied; 
\item[$\Call{TObj}{}3$]  The more positive the test objective value,  the farther  the system from violating its requirement; 
the more negative, the farther the system from satisfying its requirement. 
\end{enumerate}
These conditions ensure that we can infer using the value of $\Call{TObj}{}$ whether a test cases passes or fails, and further, $\Call{TObj}{}$ serves as a distance function, estimating how far a test is from violating model requirements, and hence, it can be used to guide generation of test cases.  
The robustness semantics of STL is an example of a semantics that satisfies those conditions~\cite{fainekos2009robustness}.
An example requirement for  \luxcasestudy\ is: 
\begin{enumerate}[leftmargin=3\parindent]
\item[SatReq] \emph{``the difference among the satellite attitude and the target attitude should not exceed 2 degrees"}. 
\end{enumerate} 
This requirement can be expressed in many languages including formal logics that predicate on signals, such as Signal Temporal Logics (STL)~\cite{maler2004monitoring} and Restricted Signals First-Order Logic (RFOL)~\cite{Socrates}.  
For example, this requirement can be expressed in STL as
\begin{align}
\LTLg_{[0, \numprint{24}h]} \left( error<2 \right) \nonumber
\end{align}
where $error$ is the difference among the satellite attitude and the target attitude, $\LTLg$ is the ``globally" STL temporal operator which is parametrized with the interval $[0, \numprint{24}h]$, i.e., the property $error<2$ should hold for the entire simulation time ($\numprint{24}h$).

We define a test objective $\Call{TObj}{}$ for this requirement as 
\begin{align}
\Call{TObj}{U, M(U)}=\underset{t \in [0, \numprint{24}h]}{min} \left( error(t)-2 \right) \nonumber
\end{align} 
This is consistent with the robustness semantics of STL~\cite{fainekos2009robustness}.
This value ensures the conditions $\Call{TObj}{}1$, $\Call{TObj}{}2$ and $\Call{TObj}{}3$ since if the property is violated, i.e., there exists a time instant $t$ such that $error(t)-2<0$, 
a negative value is returned.
In the opposite case, the property is satisfied and $\Call{TObj}{U, \mathcal{M}(U)}$ returns a non negative value.
Furthermore, the more positive the test objective value,  the farther  the system from violating its requirement; and the more negative, the farther the system from satisfying its requirement.

\vspace*{.2cm}
In our work, we use the S-Taliro tool~\cite{staliro} which implements the falsification-based testing shown in Algorithm~\ref{alg:algorithmaristeo}. S-Taliro is a well-developed, open source  research tool for falsification based-testing and has  been recently classified as  ready for industrial deployment~\cite{7741019}. It has been applied to  several realistic and industrial systems~\cite{10.1007/978-3-319-77935-5_30} and based on a recent survey on the topic~\cite{7741019} is the most mature tool for falsification of CPSs. 
Further, S-Taliro supports a range of standard search algorithms such as   Simulated Annealing, Monte Carlo~\cite{Nghiem:2010:MTF:1755952.1755983}, and gradient descent methods~\cite{abbas2014functional}.

Test objectives 
can be defined manually. Alternatively, assuming that the requirements are specified in logic languages, test objectives satisfying the three conditions we described earlier can be generated automatically. In particular, we have identified two existing tools that generate quantitative test objectives from requirements encoded in logic-based languages: Taliro~\cite{fainekos2008user} and Socrates~\cite{Socrates}.  In this paper, we use Taliro since it is  integrated into 
S-Taliro. 
To do so, we specified our requirements into Signal Temporal logic (STL)~\cite{maler2004monitoring} and used Taliro to automatically convert them into quantitative test objectives capturing degrees of satisfaction and refutation conforming to our conditions  $\Call{TObj}{}1$-$\Call{TObj}{}3$ on test objectives.

\section{\ourtool}
\label{sec:ourtool}

Algorithm~\ref{alg:algorithmaristeo} shows the approximation-refinement loop of  \ourtool. 
The algorithm relies on the following inputs:  a CI-CPS model  $\mathcal{M}$ (i.e., the model under test---MUT), the  input profile $\texttt{IP}$ of MUT,  and the maximum number of iterations $\texttt{MAX\_REF}$ that can be executed by \ourtool .  In the first iteration,  an initial surrogate model $\hat{\mathcal{M}}$ is computed such that it approximates the MUT behavior (Line~\ref{ln:abstraction}). Note that $\hat{\mathcal{M}}$ is built such that it has the same input profile as $\mathcal{M}$, i.e.,  $\hat{\mathcal{M}}$ and $\mathcal{M}$ have exactly the same inputs and outputs.
  At every iteration, the algorithm applies falsification-based testing to the surrogate model $\hat{\mathcal{M}}$ in order to find a test input $U$ violating the requirement captured by  the test objective $\Call{TObj}{}$ (Line~\ref{ln:falsification}). The number $\texttt{MAX}$ of iterations of falsification-based testing for $\hat{\mathcal{M}}$ is an internal parameter of \ourtool, and in general, can be set to a high value since executing $\hat{\mathcal{M}}$ is not expensive. 
  Once $U$ is found, the algorithm checks whether $U$ leads to a violation when it is checked on the MUT (Line~\ref{ln:conformanceCheck}). Recall from Section~\ref{sec:running} that test objectives $\Call{TObj}{}$ are defined such that a negative value indicates a requirement violation.  If so,  $U$ is returned as a 
  failure-revealing test  for  $\mathcal{M}$ (Line~\ref{ln:return}). Otherwise, $U$ is  spurious  and in the next iteration it is used to refine the surrogate model  $\hat{\mathcal{M}}$  (Line~\ref{ln:refinement}). 
  If no failure-revealing test 
 for $\mathcal{M}$  is found after $\texttt{MAX\_REF}$ iterations the algorithm stops and  a null value   is returned.

\begin{algorithm}[t]
\caption{The \ourtool\ Main Loop.}
\label{alg:algorithmaristeo}
\begin{algorithmic}[1]
\Function{\ourtool }{$\mathcal{M}, \texttt{IP}, \texttt{MAX\_REF}$}
\Repeat 
\If{$\hat{\mathcal{M}}$ is null}
\State $\hat{\mathcal{M}}$=\Call{Approximate}{$\mathcal{M}$}; \Comment{Generate a surrogate model} \label{ln:abstraction}
\Else
\State $\hat{\mathcal{M}}$=\Call{Refine}{$\hat{\mathcal{M}}$,$U$,$\mathcal{M}$}; \Comment{Refine the surrogate model} \label{ln:refinement}		
\EndIf
\State $U$=\Call{Falsification-Test}{$\hat{\mathcal{M}}, \texttt{IP}, \texttt{MAX}$}; \label{ln:falsification}
\If{\Call{TObj}{$U$, $\mathcal{M}$($U$)} $\leq 0$}  \Comment{Test $U$ finds a real violation} \label{ln:conformanceCheck}
\State	\Return{$U$}; \label{ln:return}
\EndIf
\Until{the number of executions of $\mathcal{M}$  reaches  $\texttt{MAX\_REF}$}
\State	\Return NULL; \label{ln:noinput}
\EndFunction
\end{algorithmic}
\end{algorithm}

The falsification-based testing procedure is 
described in Section~\ref{sec:running} (Algorithm~\ref{algo:fal}).  In 
Section~\ref{sec:abstractionAndRefinement}, we describe the \textsc{Approximate} method (line~\ref{ln:abstraction}), and in Section~\ref{sec:refinement}, we describe 
the \textsc{Refine} method (line~\ref{ln:refinement}).

\begin{table*}[t]
\caption{Model structure and parameter choices for developing surrogate models.}
\vspace{-0.3cm}
\label{tab:modelstructures}
\scalebox{.8}{
\begin{tabular}{p{0.01\textwidth}  p{0.17\textwidth}  p{0.87\textwidth} 
p{0.1\textwidth}}
\toprule
\multirow{35}{*}{\rotatebox[origin=c]{90}{\emph{Linear}}} & \textbf{Model Structure} & \textbf{Equation} &  \textbf{Model Type} \\
\cmidrule{2-4}
& \texttt{arx($na,nb,nk$)}
&
$y(t)=a_1 \cdot y(t-1)+\ldots+a_{na} \cdot y(t-na)+b_1 \cdot u(t-nk)+\ldots+b_{nb} \cdot u(t-nb-nk+1)+e(t)$
&
Discrete\\
\cmidrule{2-4}
& \multicolumn{3}{p{1.2\textwidth}}{\textbf{Description}}\\
\cmidrule{2-4}
& \multicolumn{3}{p{1.2\textwidth}}{
The output $y$ depends on previous input  values, i.e.,  $u(t-nk),$\ldots$,u(t-nb-nk+1)$,  and on values assumed by the output $y$  in previous steps, i.e., $y(t-1),$\ldots $,y(t-na)$. $na$ and $nb$ are the number of past output  and input values to be used in predicting the next output. $nk$ is the delay (number of samples) from the input to the output.
}
\\
 \cmidrule[\tickness]{2-4}
& \textbf{Model Structure} & \textbf{Equation} &  \textbf{Model Type} \\
\cmidrule{2-4}
& \texttt{armax($na,nb,nk,nc$)}
&
$y(t)=a_1 \cdot y(t-1)+\ldots+a_{na} \cdot y(t-na)+b_1 \cdot u(t-nk)+\ldots+b_{nb} \cdot u(t-nb-nk+1)+c_1 \cdot e(t-1)+ \ldots +c_{nc} \cdot e(t-nc)+e(t)$
&
Discrete
\\
\cmidrule{2-4}
& \multicolumn{3}{p{1.2\textwidth}}{\textbf{Description}}\\
\cmidrule{2-4}
& \multicolumn{3}{p{1.2\textwidth}}{
Extends the \texttt{arx} model by considering how the values  $e(t-1),$\ldots$,e(t-nc)$ of the noise $e$ at time $t$, $t-1$, $\ldots$,  $t-nc$ influence the value $y(t)$ of the output $y$.
}\\
 \cmidrule[\tickness]{2-4}
& \textbf{Model Structure} & \textbf{Equation} &  \textbf{Model Type} \\
\cmidrule{2-4}
& \texttt{bj($nb,nc,nf,nd,nk$)}
&
$y(t)= \frac{B(z)}{F(z)} \cdot u(t)+ \frac{C(z)}{D(z)} \cdot e(t)$
&
 Discrete
\\
\cmidrule{2-4}
& \multicolumn{3}{p{1.2\textwidth}}{\textbf{Description}}\\
\cmidrule{2-4}
& \multicolumn{3}{p{1.2\textwidth}}{
Box-Jenkins models allow a more general noise description than \texttt{armax} models.
The output $y$ depends on a finite number of previous input $u$ and output $y$ values. 
The values  $n_b$, $n_c$, $n_d$, $n_f$, $n_k$ indicate the parameters of the matrix $B$, $C$, $D$, $F$ and the value of the input delay. 
}\\
\cmidrule[\tickness]{2-4}
& \textbf{Model Structure} & \textbf{Equation} &  \textbf{Model Type} \\
\cmidrule{2-4}
& \texttt{tf($np,nz$)}
&
$y(t)= \frac{b_0+b_1 \cdot s+b_2 \cdot s^2 +\ldots + b_n \cdot s^{n_z}}{1+f_1 \cdot s + f_2 \cdot s^2+ \ldots + f_m \cdot s^{n_p}} \cdot u(t)+e(t)$
&
Continuous\\
\cmidrule{2-4}
& \multicolumn{3}{p{1.2\textwidth}}{\textbf{Description}}\\
\cmidrule{2-4}
& \multicolumn{3}{p{1.2\textwidth}}{
Represents a transfer function model. 
The values $n_p$, $n_z$ indicate the  number of poles and zeros of the transfer function.
}  \\
\cmidrule[\tickness]{2-4}
& \textbf{Model Structure} & \textbf{Equation} &  \textbf{Model Type} \\
\cmidrule{2-4}
& \texttt{ss($n$)}
&
$x(0) = x0$  \newline
$\dot{x}(t) = Fx(t) + Gu(t) + Kw(t)$ \newline
$y(t) = Hx(t) + Du(t) + w(t)$ 
&
Continuous
\\
\cmidrule{2-4}
& \multicolumn{3}{p{1.2\textwidth}}{\textbf{Description}}\\
\cmidrule{2-4}
& \multicolumn{3}{p{1.2\textwidth}}{
Uses state variables to describe a system by a set of first-order differential or difference equations.
 $n$ is an integer indicating the size of the matrix $F$, $G$, $K$, $H$ and $D$.
}\\
\toprule
\toprule
\multirow{16}{*}{\rotatebox[origin=c]{90}{\emph{Non Linear}}}  & \textbf{Model Structure} & \textbf{Equation} &  \textbf{Model Type} \\
\cmidrule{2-4}
& \texttt{nlarx($f,na,nb,nk$)}
&
$y(t) = f(y(t - 1), ..., y(t - na), u(t - nk), ..., u(t -nk -nb + 1))$
& Discrete \\
\cmidrule{2-4}
& \multicolumn{3}{p{1.2\textwidth}}{\textbf{Description}}\\
\cmidrule{2-4}
& \multicolumn{3}{p{1.2\textwidth}}{
Uses a non linear function $f$ to describe the input/output relation. 
Wavelet, sigmoid networks or neural networks in the Deep Learning Matlab Toolbox~\cite{deep} can be used to compute the function $f$.
$na$ and $nb$ are the number of past output 
and input 
values used to predict the next output value. 
$nk$ is the delay from the input to the output.}  \\
\cmidrule[\tickness]{2-4}
& \textbf{Model Structure} & \textbf{Equation} &  \textbf{Model Type} \\
\cmidrule{2-4}
& \texttt{hw($f,h,na,nb,nk$)}
&
$w(t)= f(u(t))$ \newline
$x(t)=(B(z)/F(z))\cdot w(t)$\newline
$y(t)= h(x(t))$ & 
Continuous \\
\cmidrule{2-4}
& \multicolumn{3}{p{1.2\textwidth}}{\textbf{Description}}\\
\cmidrule{2-4}
& \multicolumn{3}{p{1.2\textwidth}}{ 
Hammerstein-Wiener models describe dynamic systems two nonlinear blocks in series with a linear block. Specifically, $f$ and $h$ are non linear functions, 
$B(z)$, $F(z)$, $na$, $nb$, $nk$ are defined as for \texttt{bj} models.
Different nonlinearity estimators can be used to learn $f$ and $h$ similarly to the \texttt{nlarx} case.
}  \\
\bottomrule
\end{tabular}}
\end{table*}

\subsection{Approximation}
\label{sec:abstractionAndRefinement}
Given an MUT $\mathcal{M}$, the goal of the approximation 
is to produce a surrogate model $\hat{\mathcal{M}}$  such that: 
\begin{enumerate*}[label=(\roman*)]
\item[\textbf{(C1)}] $\mathcal{M}$ and $\hat{\mathcal{M}}$  have the same interface, i.e., the same inputs and outputs;
\item[\textbf{(C2)}] provided with the same input values, they generate similar output values; and
\item[\textbf{(C3)}] $\hat{\mathcal{M}}$ is less expensive to execute than $\mathcal{M}$
\end{enumerate*}.  

We rely on System Identification  (SI) techniques to produce surrogate models~\cite{soderstrom1989system} since their purpose is to automatically build mathematical models of dynamical systems from data when it is difficult to build the models analytically, or when engineers want to build models from data obtained based on measurements of the actual hardware. 
Note that the more complex SI structures (i.e., non-linear  \texttt{nlarx}  and  \texttt{hw}) rely on  machine learning and neural network algorithms~\cite{ljung2008system}.

To build $\hat{\mathcal{M}}$ using SI, we need some input and output data from the MUT $\mathcal{M}$. Since $\mathcal{M}$ is expensive to execute, to build the initial surrogate model $\hat{\mathcal{M}}$ (line~\ref{ln:refinement}), we run $\mathcal{M}$ for one input $U$ only. Note that an input $U$ of $\mathcal{M}$ is a set $\{u_1, \ldots, u_m\}$ of signal functions over $T=[0,b]$. So, each $u_i$ is a sequence $u_i(0), u_i(\delta), u_i(2\cdot \delta) \ldots u_i(l\cdot \delta)$ where $b =  l\cdot \delta$ and $\delta$ is the sampling rate applied to the time domain $[0,b]$. Similarly, the output $Y = \mathcal{M}(U)$ is a set $\{y_1, \ldots, y_n\}$ of signal functions where each $y_j$ is a sequence $y_j(0), y_j(\delta), y_j(2\cdot \delta) \ldots y_j(l\cdot \delta)$ obtained based on the same sampling rate and the same time domain as those used for  the input.  We refer to the data used to build  $\hat{\mathcal{M}}$ as \emph{traning data} and denote it by $\mathcal{D}$. Specifically,  
$\mathcal{D}=\langle U, Y \rangle$.
For CI-CPS, the size $l$ of  $\mathcal{D}$
tends to be large since we typically execute such models for a long time duration (large $b$) and use a small sampling rate (small $\delta$) for them. For example, we typically run \luxcasestudy\ for $b = \numprint{86400}$s ($24$h) and use the sampling rate $\delta = 0.0312$s. Hence,  a single execution of \luxcasestudy\ generates a training data set $\mathcal{D}$ with size $l=\numprint{2769200}$.   
Such training data size is sufficient for SI to build reasonably accurate surrogate models.

We use the System Identification Toolbox~\cite{ljung2008system} of Matlab to generate surrogate models. In order to effectively use SI, we need to anticipate the expected \emph{structure} and \emph{parameters} of  surrogate models, a.k.a \emph{configuration}. Table~\ref{tab:modelstructures} shows some standard model  structures  and parameters supported by SI. 
Specifically, selecting the model structure is about deciding which  mathematical equation among those shown in Table~\ref{tab:modelstructures} is more likely to fit to our training data and is better able to capture the dynamics  of the  model $\mathcal{M}$. As shown in Table~\ref{tab:modelstructures}, equations specifying the model structure have some parameters that need to be specified so that we can apply SI techniques. For example, for \texttt{arx($na,nb,nk$)}, the values of the parameters $na$, $nb$ and $nk$ are the model parameters.

Table~\ref{tab:modelstructures}  provides a short description for each model structure. 
We note that some of  the equations in the table are simplified and refer to the case in which the MUT has a single input signal
and a single output signal. 
The equations, however, can be generalized to models with 
multiple input and output signals.  Briefly, model structures can be linear or non-linear in terms of the relation between the  inputs and outputs, or they can be continuous and discrete in terms of their underlying training data. Specifically, the training data generated from MUT can be either discrete (i.e., sampled at a fixed rate) or continuous  (i.e., sampled at a variable rate). Provided with discrete training data, we can select either continuous or discrete model structures, while for continuous training data, we can select continuous model structures only. As discussed earlier, our training data $\mathcal{D}$ is discrete since it is sampled at the fix sampling rate of $\delta$. Hence, we can choose both types of model structures to generate surrogate models. 
In our work  we 
support training data sampled at a fixed sampling rate to build and refine the surrogate models. 
Data sampled at a variable time rate can be then handled by exploiting the resampling procedure of Matlab~\cite{Resampling}.

The users of \ourtool\ 
 need to choose upfront the configuration to be used by the SI, i.e., the model structure and the values of its parameters. 
This choice  depends on domain specific knowledge that the engineers possess for the model under analysis. 
The values of the parameters selected by the user should be chosen such that  the resulting surrogate model (i) has the same interface as the MUT  to ensure \textbf{C1} and  (ii)  has a simpler structure than the MUT  to ensure \textbf{C3}.
The System Identification Toolbox provides some generic guidance for selecting the parameters ensuring these two criteria~\cite{modelStructure}.
In this work we performed  an empirical evaluation  over a set of benchmark models  to determine the configuration to be used in our experiments (Section~\ref{sec:modelstructure}).

Once a configuration 
is selected, SI uses the training data to learn values for the coefficients of the equation from  Table~\ref{tab:modelstructures} that corresponds to the selected structure and paramters. For example, after selecting \texttt{arx($na,nb,nk$)} and assigning values to $na$, $nb$ and $nk$,  SI generates a surrogate model by learning values for the coefficients: $a_1,\ldots a_{na}$ and $b_1,\ldots b_{nb}$.

Similar to standard machine learning algorithms,  SI's objective is to compute the model coefficients by  minimizing the difference (error) between the outputs of $\mathcal{M}$ and $\hat{\mathcal{M}}$ for the training data~\cite{soderstrom1989system}.  SI uses  different standard notions of errors depending on the model structure selected. In our work, we compute the  Mean Squared Error (MSE)~\cite{soderstrom1989system}  between the outputs of $\mathcal{M}$ and $\hat{\mathcal{M}}$.

SI 
learns a surrogate model $\hat{\mathcal{M}}$ by minimizing MSE over the training data $\mathcal{D}$ and hence, ensuring \textbf{C2}. The learning algorithm selected by SI depends on the chosen model structure, on the purpose of the identification process, i.e., whether the identified model will be used for prediction or simulation, and on whether the system is continuous or  discrete.

\subsection{Refinement}
\label{sec:refinement}
The refinement step rebuilds the surrogate model  $\hat{\mathcal{M}}$ when the test input $U$ obtained by falsification-based testing of the surrogate model is 
spurious for MUT (i.e., it does not reveal any failure according to the test objective). Note that $\hat{\mathcal{M}}$ may not be sufficiently accurate to predict the behavior of the MUT. Hence, it is likely that we need to improve its accuracy and we do so by reusing the data obtained when  checking  a candidate test input $U$ on MUT (line~\ref{ln:conformanceCheck} of  Algorithm~\ref{alg:algorithmaristeo}).  

Let $U=\{u_1, \ldots, u_m\}$ and $Y=\{y_1, \ldots, y_n\}$ be the spurious test inputs and its output, respectively. 
Similar to the data used to build the initial  $\hat{\mathcal{M}}$  by the approximate step (line~\ref{ln:conformanceCheck} of  Algorithm~\ref{alg:algorithmaristeo}), 
the data $\mathcal{D}'=\langle U,Y \rangle$ used to rebuild $\hat{\mathcal{M}}$  is also discretized based on the same sampling rate $\delta$. 
To refine the surrogate model, we  do not 
change 
the considered configuration,  
but we combine the new 
 $\mathcal{D}'$ 
 and existing training data  $\mathcal{D}$, and refine $\hat{\mathcal{M}}$ using these data.

Alternative policies can be chosen to refine the surrogate model.
For example, the refinement activity may also change the configuration of \ourtool . 
This is 
 a rather drastic change in the surrogate model. 
 When engineers have a 
 clear understanding of the underlying model, they may be able to define a systematic methodology on how to move from less complex structures (e.g., linear) to more complex ones (e.g., non-linear). Without proper domain knowledge, such modification may 
 be too disruptive. 
 In this paper, our refinement strategy is focused on incrementing the training data and rebuilding the surrogate model without changing the 
  configuration.

\section{Evaluation}
\label{sec:evaluation}
In this section, we empirically evaluate \ourtool\ by answering the following research questions: 

\noindent $\bullet$ \textbf{Configuration - RQ1.} \emph{Which are the optimal (most effective and efficient) SI configurations for \ourtool ? Which of the optimal configurations can be used in the rest of our experiments?} 
We investigate the performance of \ourtool\  for different SI configurations (model structures and parameters listed in Table~\ref{tab:modelstructures}) to identify the optimal ones, i.e., those that offer the best trade-offs between effectiveness (revealing the most requirements violations) and efficiency (revealing the violations in less time).  We then select one configuration among the optimal ones and use that configuration for the rest of our experiments.

\noindent $\bullet$ \textbf{Effectiveness - RQ2.} \emph{How \emph{effective} is \ourtool\ in generating tests that reveal requirements violations?} We use \ourtool\ with the optimal configuration identified in {\bf RQ1} and evaluate its effectiveness (i.e., its ability in detecting requirements violations) by comparing it with  falsification-based testing without surrogate models. We use S-Taliro discussed in Section~\ref{sec:running} for the baseline of comparison. 

\noindent $\bullet$ \textbf{Efficiency - RQ3.} \emph{How \emph{efficient} is \ourtool\  in generating tests revealing requirements violations?} We use \ourtool\ with the optimal configuration identified in {\bf RQ1} and evaluate its efficiency  (i.e., the time it takes to find violations) by comparing it with falsification-based testing without surrogate models (i.e., S-Taliro).

A key challenge regarding the empirical evaluation of  \ourtool\  is that, both \ourtool\   and S-Taliro rely on randomized algorithms. Hence, we have to repeat our experiments numerous times for different models and requirements so that the results can analysed in a sound and systematic way using statistical tests~\cite{Hitchhiker}.  This is  necessary to  answer RQ1-RQ3 that involve selecting an optimal configuration  and comparing \ourtool\  with the baseline  S-Taliro. Performing these experiments on CI-CPS models is, however, extremely expensive, to the point that the experiments become infeasible. A ballpark figure for the execution time of the experiments required to answer RQ1-RQ3 is around 50 years if the experiments are performed on our CI-CPS model case study (\luxcasestudy). Therefore, instead of using CI-CPS models, we use non-CI-CPS models to address RQ1-RQ3. The implications of this decision on the results are assessed and mitigated in Sections~\ref{sec:modelstructure} and~\ref{subsec:rq2-rq3} where we discuss these three research questions in detail. 
In addition, to be able to still assess the performance of \ourtool\ on CI-CPS models, we consider an additional research question described below:

\begin{table}[t]
\caption{Non-CI-CPS subject models. 
\textbf{ID}: model identifier;
\textbf{\#B}: number of  blocks of the Simulink model;
\textbf{\#I}: number of inputs of the Simulink model; 
\textbf{int(n)}: input interpolation functions and number of control points;
\textbf{R}: input ranges; and
\textbf{T}: time domain.
} 
\vspace{-0.3cm}
\label{table:benchmarkmodels}
\scalebox{.8}{\begin{tabular}{ c  c c   c   c   c }
\toprule
\bf ID &  \bf \#B &  \bf \#I & \bf int(n) & \bf R &    \bf T \\
\toprule
RHB(1) & 28 & 1 & \texttt{pchip(4)}  & $[-2, 5]^*$  & 24  \\
\midrule
RHB(2) & 31 & 2 &\makecell{\texttt{pchip(4)},\texttt{const(1)}} & \makecell{$[-2, 5]$, $[0.8, 1.2]$}&  24  \\
    \midrule
AT& 63 & 1 &\makecell{\texttt{pconst(7)}} & $[0,100]$  & 30 \\
\midrule
AFC &  302 & 2 &\makecell{\texttt{const(1)},\texttt{pulse(10)}} & 
\makecell{$ [900,1100] $, $[0,61.1]$}  &  50\\
  \midrule
IGC & 70  & 10 & 
\makecell{\texttt{const(1)},\texttt{const(1)},\\
\texttt{const(1)},\texttt{const(1)},\\
\texttt{const(1)},\texttt{const(1)},\\
\texttt{const(1)},\texttt{const(1)},\\
\texttt{const(1)},\texttt{const(1)}}
 & 
\makecell{$[40,40]$,$[30,30]$,\\
$[200,200]$,$[40,40]$,\\
$[150,250]$,$[0,80]$,\\
$[20,50]$,$[100,300]$,\\
$[20,70]$,$[-0.3,0.3]$} & 400    \\
\bottomrule
\end{tabular}}
\end{table}

\noindent $\bullet$ \textbf{Usefulness - RQ4.} \emph{How applicable and useful is \ourtool\ in generating tests revealing requirements violations for industrial CI-CPS models?} We apply  \ourtool\  with the optimal configuration identified in {\bf RQ1}  to our  CI-CPS model case study from the satellite industry (\luxcasestudy) and evaluate its effectiveness and efficiency. The focus here is to obtain representative results in terms of effectiveness and efficiency based on an industry CI-CPS model. Note that we still apply S-Taliro to \luxcasestudy\ to be able to compare it with \ourtool\ for an industry CI-CPS model. This comparison, however, is not meant to be subject to statistical analysis  due to the large execution time of \luxcasestudy, and is only meant to complement  RQ3 with a fully realistic though extremely time consuming study.

\textbf{The subject models.} We used five publicly available non-CI-CPS models (i.e., RHB(1), RHB(2), AT, AFC, IGC)
that have been previously used in the literature on falsification-based testing of CPS models~\cite{fehnker2004benchmarks,dang2004verification,zhao2003generating,jin2014powertrain,sankaranarayanan2012simulating,ernst2019arch}.
The models  represent realistic and representative models of CPS systems from different domains. 
RHB(1) and RHB(2)~\cite{fehnker2004benchmarks} are from the IoT and smart home domain.
AFC~\cite{jin2014powertrain} is from the automotive domain and has been originally developed by Toyota. 
AT~\cite{zhao2003generating} is another model  from the automotive domain. 
IGC~\cite{sankaranarayanan2012simulating} is from the health care domain. 
AT and AFC have also been recently considered as a part of the reference benchmarks in the ARCH competition~\cite{ernst2019arch} -- an 
international competition among verification and testing tools for continuous and hybrid systems~\cite{cpsweek}. 
 In Table~\ref{table:benchmarkmodels}, we report the number of blocks and inputs, the input profiles, input ranges and simulation times for the five non-CI-CPS models.
These models have been manually developed  and  may violate their requirements due to human error.  Some of the violations have been identified by the existing testing tools and are reported in the literature~\cite{fehnker2004benchmarks,jin2014powertrain,zhao2003generating,sankaranarayanan2012simulating,ernst2019arch}. 
As for the CI-CPS model to address RQ4, we use the \luxcasestudy\ case study  that we introduced as a running example in Sections~\ref{sec:running} and \ref{sec:ourtool}. 
 \luxcasestudy\ contains $2192$  blocks  and has to be simulated for $24$h for each test case to sufficiently exercise the system dynamics and interactions with the environment.  Like the models in Table~\ref{table:benchmarkmodels}, \luxcasestudy\ is manually developed by engineers and is likely to be faulty. Its inputs and input profiles are  shown in Table~\ref{table:rq3}.

\textbf{Implementation and Data Availability.} We implemented \ourtool\ as a Matlab  application and as an add-on of  S-Taliro. 
Our (sanitized) models, data and tool are available online~\cite{appedix} and are also submitted alongside the paper.

\subsection{RQ1 - Configuration}
\label{sec:modelstructure}
Recall that  \ourtool\ requires to be provided with a configuration to build surrogate models.
The universe of the possible configurations is infinite as the model structures in Table~\ref{tab:modelstructures} can be parametrized in an infinite number of ways by associating different values to their parameters.    
RQ1 identifies the optimal configurations that yield the best tradeoff between effectiveness and efficiency for  \ourtool\ among a reasonably large set of 
alternative representative configurations.
It then selects one among the optimal configurations.

We do not evaluate configurations by measuring their prediction accuracy (i.e., by measuring their prediction error when applied to a set of test data as is common practice in assessing prediction models in the machine learning area~\cite{bishop2006pattern}) 
because our focus is not to have the most accurate configuration but the one that is able to have the most effective impact on \ourtool's approximation-refinement loop by quickly finding  requirements violations. 
However, it is likely that there exists a relationship between the two.  

\textbf{Experiment design.} 
We consider five different configurations obtained by five different sets of parameter values for  each model structure in Table~\ref{tab:modelstructures}.   We denote the five configurations related to each model  structure $S$ by $S_1$ to $S_5$. For example, the configurations related to the model structure \texttt{ss} are denoted by  \texttt{ss}$_1$ to  \texttt{ss}$_5$.   The specific parameter value sets for the 35 configurations based on the seven model structures in Table~\ref{tab:modelstructures} are available online~\cite{appedix}. 

To answer RQ1, we apply \ourtool\  to the five non-CI-CPS models  using each configuration among  the 35 possible ones.  That is, we execute \ourtool\  for 175 times. We further rerun each application of \ourtool\  for 100 times to account for the randomness in both falsification-based testing and the approximation-refinement loop  of \ourtool~\cite{8453180}. 
We set the value of \texttt{MAX\_REF}, i.e, the number of iterations of the  \ourtool's main loop,  to 10 (see Algorithm~\ref{alg:algorithmaristeo}) and the value of \texttt{MAX}, i.e, the number times each iteration of  \ourtool\ executes falsification-based testing (see Algorithm~\ref{algo:fal}), to $100$ for RHB(1), RHB(2) and AFC, and to $1000$ for AT and IGC. These values were used 
in the original experiments that apply falsification-based testing to these models~\cite{benchmarkconfigurations}.  Running all the 17,500 experiments required 
 \totaltimerunningrqone\ hours ($\approx$ \totaltimerunningrqonedays\ days).\footnote{
We used the high performance computing cluster at [location redacted]
with $100$ Dell PowerEdge C6320 and a total of 2800 cores with 12.8 TB RAM.
The parallelization  reduced the experiments time to approximately $15$ days.}

\begin{figure}[t]
\centering
\includegraphics[width=\columnwidth]{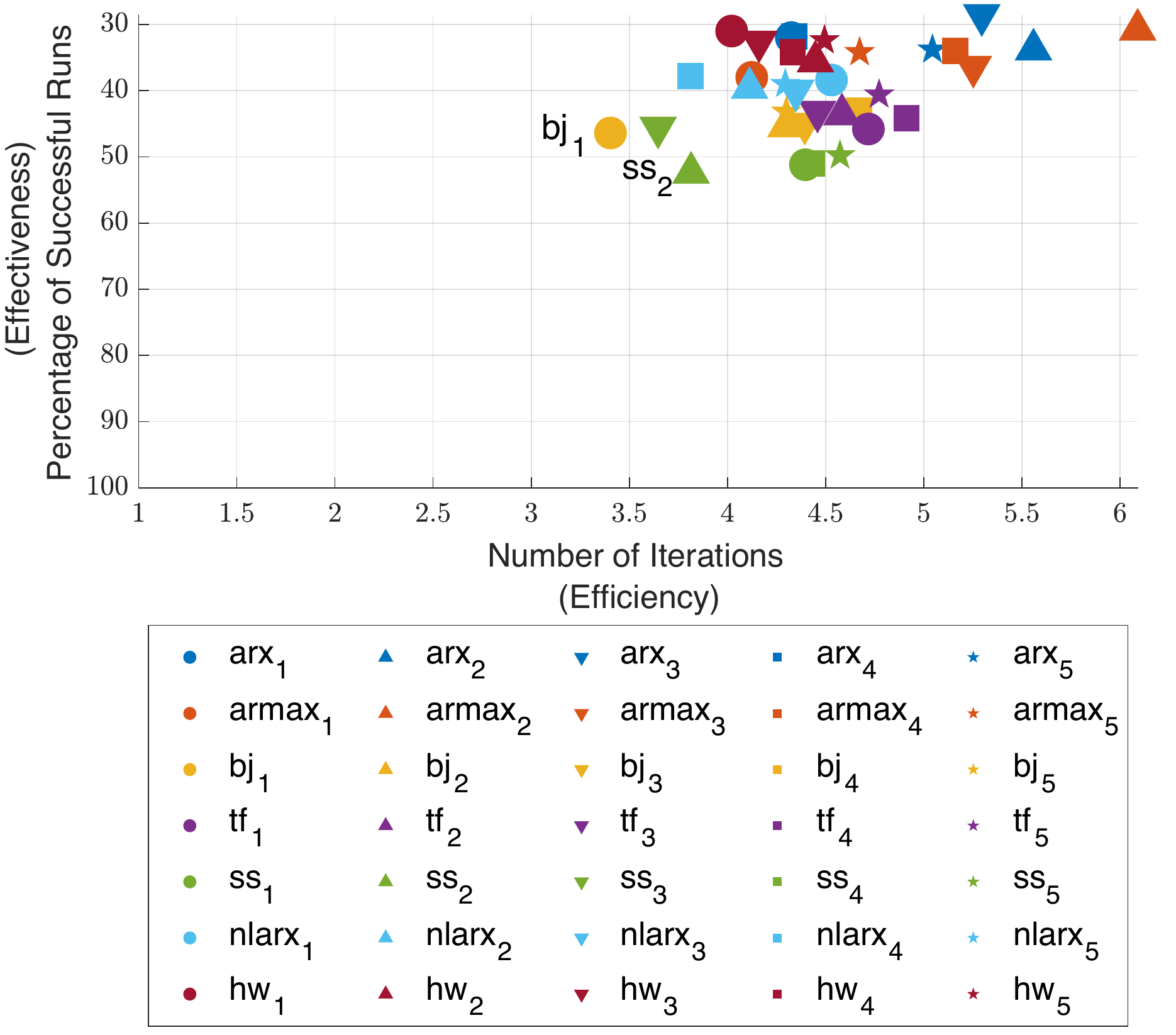}
\vspace{-0.7cm}
\caption{Effectiveness and efficiency of different configurations across our non-CI-CPS subject models.}
\label{fig:RQ1}
\end{figure}

Due the sheer size of the experiments required to answer RQ1, we used  our non-CI-CPS subject models. 
While these models are smaller than typical CI-CPS models, 
the complexity of their  structure (how Simulink blocks are used and connected) is similar to the one of \luxcasestudy.
Specifically, the structural complexity index~\cite{plkaska2009quality,olszewska2011simulink}, which provides an estimation of the complexity of the structure of a Simulink model, 
is $1.8$, $1.6$, $1.2$, $1.1$, $2.1$ for the RHB(1), RHB(2), AT, AFC and IGC benchmarks, respectively, and  $1.5$ for the \luxcasestudy\ case study.
 We conjecture that given these similarities, the efficiency and effectiveness comparisons of the configurations performed on non-CI-CPS models would likely remain the same should the comparisons be performed on CI-CPS models. However, due to computational time restrictions, we are not able to check this conjecture.  Finally, we note that even if we select a sub-optimal configuration, it will be a disadvantage for  \ourtool. So, the results for RQ2-RQ4  are likely to improve if we find a way to identify a better configuration for \ourtool\ using CI-CPS models.

\textbf{Results.} 
The scatter plot in Figure~\ref{fig:RQ1}  shows the results of our experiments. 
The x-axis 
indicates our \emph{efficiency metric} which is defined as \emph{the number of iterations that  \ourtool\ requires to reveal a requirement violation} in a model for a given configuration.  
As described in the experiment design, the maximum number of iterations is 10. Given a configuration for  \ourtool, the fewer iterations required to reveal a violation, the more efficient that configuration is. 
The y-axis 
 indicates our \emph{effectiveness metric} which is defined as \emph{the number of \ourtool\  runs (out of 100) that can reveal a violation} in a model. 
 For  effectiveness 
 we are interested to know how often we are able to reveal a requirement violation. 
 The higher the number of runs detecting violations, the more effective that configuration is. The ideal configuration is the one that 
 finds requirements violations in  100\% of the runs in just one iteration as indicated by the origin 
 of the plot in Figure~\ref{fig:RQ1} with coordinates $(1, 100)$.

 For each configuration, there is one point in the plot in Figure~\ref{fig:RQ1} whose coordinates, respectively, indicate the average efficiency and effectiveness of that configuration for the non-CI-CPS subject models. As shown in the figure, 
 bj$_1$ and ss$_2$ are on the Pareto frontier~\cite{pareto} and  dominate other configurations in terms of efficiency and effectiveness. That is, any configuration other than  bj$_1$ and ss$_2$ is strictly dominated in terms of both efficiency and effectiveness by either   bj$_1$ or ss$_2$.
But bj$_1$ does not dominate  ss$_2$, and neither does ss$_2$. Specifically, bj$_1$ is more efficient but less effective than  ss$_2$, and  ss$_2$ is less efficient but more effective than bj$_1$.  For our experiments, we select bj$_1$ as the optimal configuration 
since efficiency is paramount when  dealing with CI-CPS models. In terms of effectiveness, bj$_1$ is only slightly less effective than  ss$_2$ (46.4\% versus 52.4\%).

\resq{The answer to \textbf{RQ1} is that, among all the 35 configurations we compared, the bj$_1$ and the ss$_2$ configurations are the optimal configurations offering the best trade-off between efficiency (i.e., time required to reveal requirements violations) and effectiveness (i.e., number of violations revealed) for \ourtool. We select bj$_1$ as we prioritize efficiency.}

\subsection{RQ2 and RQ3 - Effectiveness and Efficiency}
\label{subsec:rq2-rq3}
For RQ2 and RQ3, we compare \ourtool\ (Algorithm~\ref{alg:algorithmaristeo}) with S-Taliro (Algorithm~\ref{algo:fal}).  As discussed earlier, due to the large size of the experiments, we  use non-CI-CPS models, \emph{but we want to obtain results that are representative for the CI-CPS case}. For such comparisons, we need to execute both tools for an equivalent amount of time and then compare  their effectiveness and efficiency. 
This is a non trivial problem, because
\begin{itemize}
\item That equivalent amount of time cannot simply translate into identical \emph{execution times}. Non-CI-CPS models, by definition, are  very quick to execute. Hence, the benefits of performing the falsification on the surrogate model, as done by \ourtool ,   would not be visible if we compared the two tools  
based on the execution times of non-CI-CPS models. Therefore, comparisons would be in favour of S-Taliro if we fix the execution times of the two tools for non-CI-CPS models. 
\item Neither can we can  run the two tools for the same \emph{number of iterations}, as commonly done in this domain~\cite{ernst2019arch}, because one iteration of  \ourtool\ takes more time  than one iteration of S-Taliro. Recall that  \ourtool, in addition to performing falsification,  builds and refines surrogate models in each iteration. Thus, by fixing the number of iterations for the two tools,  comparisons would be in favour of \ourtool. 
\end{itemize}

To  answer RQ2 and RQ3 without favouring neither of the tools, we propose the following:

Suppose that we could perform {\bf RQ2} and {\bf RQ3} on  a  CI-CPS model, and that we execute  \ourtool\ and S-Taliro on this model for the same time limit $TL$. Let   $IA$ and $IB$ be the number of iterations of   \ourtool\ and S-Taliro within $TL$, respectively. Recall that one iteration of  \ourtool\  typically takes more time than one iteration of the baseline ($IA < IB$). If we know the values of  $IA$ and $IB$, we can execute \ourtool\ for $IA$ times and S-Taliro for $IB$ times on non-CI-CPS models and use the results to compare the tools as if they were executing on CI-CPS models.

To run our experiment, 
we need to know the relation between  $IA$ and $IB$. We approximate this relation empirically using our \luxcasestudy\   CI-CPS model.  We execute \ourtool\ for $10$ iterations and we set the number of falsification iterations in each iteration of \ourtool\  to $100$ as suggested by the literature on CPS falsification testing~\cite{staliro,benchmarkconfigurations} (i.e.,  \texttt{MAX\_REF}  = 10 and \texttt{MAX} = $100$ in Algorithm~\ref{alg:algorithmaristeo}). We repeated these runs of   \ourtool\  for five times. The first iteration of  \ourtool\  took, on average, \numprint{16902}s, and the subsequent iterations of \ourtool\  took, on average,  \numprint{9865}s. Note that the first iteration of \ourtool\   is always more expensive than the subsequent iterations since  \ourtool\ builds surrogate models in the first iteration. Similarly, we executed S-Taliro for 10 iterations on \luxcasestudy, and repeated this run for five times. Each iteration of S-Taliro took, on average, \numprint{8336}s on  \luxcasestudy. 
This preliminary experiment took approximately 20 days. 
 We then solve the two equations  below to approximate the relation between  $IA$ and $IB$:
\begin{align}
& TL = \numprint{9865} \times (IA-1) +\numprint{16902} &&\label{eq1}\\[-3pt]
& TL = \numprint{8336} \times IB&& \label{eq2} 
\end{align}
 The above yields $IB = 1.2 \times IA + 0.8$. Though we obtained this relation between $IA$ and $IB$ based on one CI-CPS case study,   \luxcasestudy\  is a large and industrial system representative of the CPS domain.  Further, for CI-CPS models that are more compute-intensive than \luxcasestudy, executing the models takes even more time compared to the approximation and refinement time, and hence, the relation above could be further improved in favour of  \ourtool.

\begin{figure*}[ht]
\includegraphics[width=\textwidth]{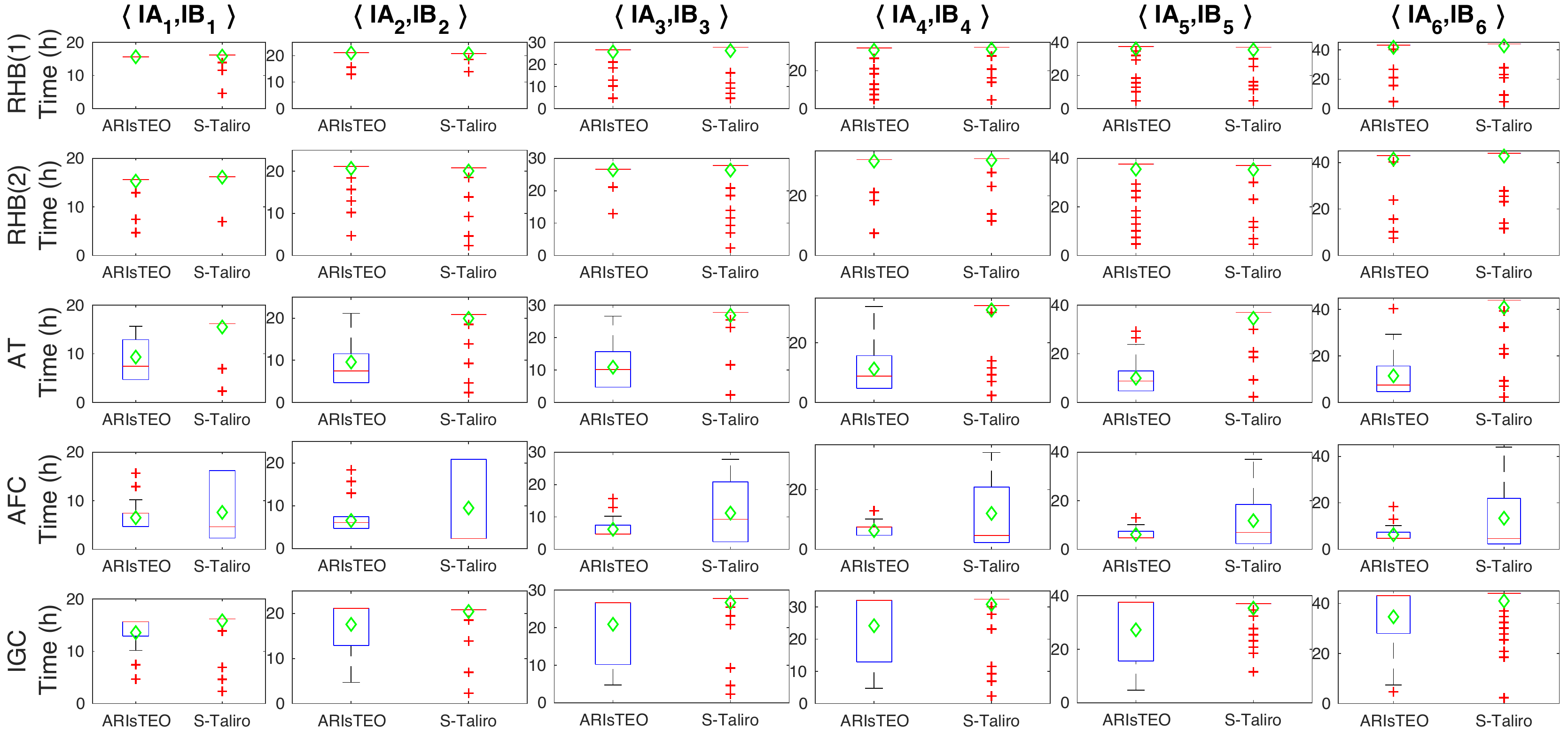}
\vspace{-0.6cm}
\caption{Comparing the efficiency of \ourtool\ and  S-Taliro.
The box plots show the execution time (computed using equations~\ref{eq1} and~\ref{eq2}) of \ourtool\ and S-Taliro (in hours) for our non-CI-CPS subject models (labels on the left of the figure)
and over different iterations (labels on the top of the figure). Diamonds depict the average.}
\label{fig:efficiency}
\end{figure*}

\begin{table}[t]
\caption{The effectiveness results.
Percentages of cases in which  \ourtool\ (IA$_i$ labelled columns) and S-Taliro (IB$_i$ labelled columns) were able to detect requirements violations for different iteration pairs (IA$_i$ and IB$_i$) and  benchmarks.}
\vspace{-0.3cm}
\label{table:rq1results}
\scalebox{.68}{
\begin{tabular}{c | c c | c c | c c | c c | c c | c c }
\toprule
 &   IA$_1$ & IB$_1$ &  IA$_2$ & IB$_2$ &  IA$_3$ & IB$_3$ &  IA$_4$ & IB$_4$ &  IA$_5$ & IB$_5$ &  IA$_6$ & IB$_6$ \\ 
\midrule
RHB(1) 		& 0\% 		& 5\% 	& 2\% 		& 2\% 	& 8\% 		& 8\%  	& 7\% & 7\% & 11\% & 6\% & 9\% & 8\% \\ 
RHB(2) 		& 5\% 		& 2\% 	& 6\% 		& 8\% 	& 4\% 		& 9\% 	& 5\% & 10\% & 13\% & 10\% & 7\% & 10\% \\ 
AT 			& 85\% 		& 7\% 	& 92\% 		& 7\% 	& 93\% 		& 7\% 	& 99\% & 4\% & 100\% & 8\% & 100\% & 13\% \\ 
AFC 			& 100\%	 & 77\%	 & 100\% 	& 73\% & 100\% 	& 88\% & 100\% & 86\% & 100\% & 92\% & 100\% & 95\% \\ 
IGC 			& 33\% 		& 4\% 	 & 31\% 	& 6\% 	& 34\% 		& 9\% & 37\% & 15\% & 40\% & 18\% & 13\% & 21\% \\ 
\bottomrule
\end{tabular}}
\vspace{-0.3cm}
\end{table}

\textbf{Experiment design.} 
To answer \textbf{RQ2} and \textbf{RQ3}, we applied \ourtool\ with the configuration identified by \textbf{RQ1} (bj$_1$) and S-Taliro to the five non-CI-CPS models in Table~\ref{table:benchmarkmodels}.  We executed  \ourtool\ and S-Taliro  for the following pairs of iterations: $\langle IA_1=5 , IB_1 =7 \rangle$, $\langle IA_2=7, IB_2 =9 \rangle$, $\langle IA_3=9, IB_3 =12 \rangle$, $\langle IA_4=11, IB_4 =14 \rangle$, $\langle IA_5=13, IB_5 =16 \rangle$, and $\langle IA_6=15, IB_6 =19 \rangle$.  Note that every pair approximately satisfies $IB_i = 1.2 \times IA_i + 0.8$.  We repeated each run  100 times to account for their randomness.  
For \textbf{RQ2}, we compute the \emph{effectiveness metric} as  in {\bf RQ1}: the number of runs revealing requirements violations (out of 100) for each tool. 
For \textbf{RQ3}, we assess efficiency by computing the  \emph{efficiency metric} as in {\bf RQ1}: the number of iterations that  each tool requires to reveal a requirement violation. However, as discussed above, the number of iterations of 
 \ourtool\ and S-Taliro are not comparable. Hence, for \textbf{RQ3}, we report efficiency in terms of  the estimated time that each tool needs to perform those iterations on CI-CPS models computed using equations~\ref{eq1} and~\ref{eq2}.

\textbf{Results-RQ2.}  
Table~\ref{table:rq1results} shows  the effectiveness values for \ourtool\  and S-Taliro for the five iteration pairs discussed in the 
experiment design.  For the AT, AFC and IGC models, the average effectiveness of \ourtool\  is significantly higher 
than that of S-Taliro ($75.4\%$ versus $35.0\%$ on average across benchmarks), while for RHB(1) and RHB(2), \ourtool\  and S-Taliro  
reveal almost the same number of violations  ($6.4\%$ versus $7.0\%$ on average across benchmarks). 
The former difference in proportion is statistically significant as confirmed by a two-sample z-test~\cite{mcdonald2009handbook} with the level of significance ($\alpha$) set to 0.05.

RHB(1) and RHB(2) have more outputs than the other benchmarks and they have 
shorter simulation times (see Table~\ref{table:benchmarkmodels}). This is an increased challenge for building accurate surrogate models. In practice, CI-CPS models can have a large number of outputs but they usually involve long simulation times.

\resq{The answer to \textbf{RQ2} is that  \ourtool\ is significantly more effective than S-Taliro for three benchmark models while, for the other two models, they reveal almost the same number of violations. 
On average, over the five models, \ourtool\  detects $23.9\%$  more requirements violations than S-Taliro (min=-8\%, max=95\%).
}

\textbf{Results-RQ3.} 
The execution times (computed using equations~\ref{eq1} and~\ref{eq2}) of \ourtool\ and S-Taliro for our non-CI-CPS subject models 
and the iteration pairs  $\langle IA_i, IB_i \rangle$  are shown in  Figure~\ref{fig:efficiency}.
The box plots in the same row are related to the same benchmark model, while the box plots in the same column are related to the same iteration pair. Recall that we described the iteration pairs $\langle IA_i, IB_i \rangle$ considered for our experiments  earlier in the experiment design subsection. 
As expected, 
the average execution times of the two tools increases with their number of iterations.

To statistically compare the results,  we used the Wilcoxon  rank sum test~\cite{mcdonald2009handbook}  with the level
of significance ($\alpha$) set to $0.05$. The  results show that 
\ourtool\  is significantly more efficient than  S-Taliro  for the AT and IGC models (Figure~\ref{fig:efficiency} -- rows 3,5).  The efficiency improvement that 
\ourtool\  brings about over S-Taliro for AT and IGC across different iterations ranges from $14.4\%$ ($2.2$h) to $73.1\%$ ($31.2$h). Note that, for  AT and IGC,  \ourtool\ is significantly more effective than S-Taliro (see  Table~\ref{table:rq1results}). This shows that,  many runs of \ourtool\ for AT and IGC can reveal a requirement violation and stop before reaching the maximum ten iterations, hence yielding better efficiency results of \ourtool\  compared to the other model.

For the RHB(1)  and RHB(2) models  (Figure~\ref{fig:efficiency} -- rows 1,2),  \ourtool\ and S-Taliro yield  comparable  efficiency results.  
The effectiveness results in Table~\ref{table:rq1results} confirm that, for  RHB(1)  and RHB(2),  both  \ourtool\ and S-Taliro  have to execute for ten iterations most of the times as they cannot reveal violations (low effectiveness). Hence, the efficiency results are worse for RHB(1)  and RHB(2) than  for the other models. Further, as we run the tools for more iterations, the efficiency results slightly increases as indicated by the increase in the number of outliers.  For the AFC model (Figure~\ref{fig:efficiency} -- row 4), \ourtool\ is slightly  more efficient than S-Taliro. For AFC, S-Taliro is relatively effective in finding violations, and hence, is efficient.  But,  its average execution time is  slightly worse than that of ARIsTEO. Comparing the interquartile ranges of the box plots shows that ARIsTEO is generally more efficient that S-Taliro. However, a Wilcoxon test does not reject the null hypothesis (p-value = 0.06).

The average execution time of \ourtool\ and S-Taliro across the different models is, respectively, approximately $19$h and $25$h.
Though there is significant variation across the different models, \ourtool\ is, on average, $31.3\%$ more efficient than S-Taliro.

\resq{The answer to \textbf{RQ3} is that  \ourtool\ is on average $31.3\%$ (min=$-1.6\%$, max=$85.2\%$) more efficient than S-Taliro. 
}

\subsection{RQ4 - Practical Usefulness}
We assess the usefulness of \ourtool\  in revealing requirements violations of a  representative industrial CI-CPS model.

\textbf{Experiment design.} We received three different requirements from our industry partner~\cite{Luxspace}. One is the SatReq requirement  presented in Section~\ref{sec:running}, and the two others  (SatReq1 and SatReq2) are strengthened  versions of  SatReq  that, if violated, indicate increasingly critical violations.  We also received the input profile IP (Section~\ref{sec:running}) and a more restricted input profile IP$^\prime$, 
representing  realistic input subranges associated with more critical violations.  For each combination of the requirement (SatReq, SatReq1 and SatReq2) and the input profiles IP and IP$^\prime$, we checked whether \ourtool\ was able to detect any requirement violation, and further, we recorded the time needed by \ourtool\  to detect a violation.  In addition, for the two most critical requirements (SatReq1 and SatReq2) and the input profiles IP and IP$^\prime$, we checked whether S-Taliro is able to detect any violation within the time limit required by  \ourtool\ to successfully reveal  violations for SatReq1 and SatReq2.  Running this experiment took approximately four days and both tools were run twice for each requirement and input profile combination.   

\textbf{Results}.  \ourtool\ found a violation for every requirement and input profile combination in our study in just one iteration, requiring approximately four hours of execution time. Given that simulating the model under test takes approximately an hour and a half, detecting errors in four hours is highly efficient as it corresponds to roughly two model simulations. In comparison, S-Taliro  failed to find any violations for SatReq1 and SatReq2 after running the tool  for four hours based on  the input profiles  IP and IP$^\prime$. 

\vspace{0.1cm}
\resq{The answer to \textbf{RQ4} is that  
\ourtool\ efficiently detected requirements violations -- in practical time -- that S-Taliro could not find, for 
 three different requirements and two input profiles on an industrial CI-CPS model.
}

\section{Related Work}
\label{sec:related}

Formal verification techniques such as model checking aim to exhaustively check correctness of behavioural/functional models (e.g.,~\cite{fan2017d,henzinger1997hytech}), but they often face scalability issues for complex CPS models. 
The CEGAR framework has been proposed to help model checking scale to such models (e.g.,~\cite{alur2003counter,ratschan2007safety,alur2006predicate,clarke2003verification,clarke2003abstraction,
ratschan2007safety,segelken2007abstraction,HARE,
dierks2007automatic,jha2007reachability,prabhakar2015hybrid,10.1007/978-3-319-65765-3_7,sorea2004lazy,zhang2017abstraction,Nellen2016}).
As discussed in Section~\ref{sec:intro},  the approximation-refinement loop of \ourtool, at a general level, is inspired by CEGAR. 
Two CEGAR-based model checking approaches have been proposed for hybrid systems capturing CPS models:  (a)~abstracting hybrid system models into discrete finite state machines without dynamics~\cite{alur2006predicate,clarke2003verification,clarke2003abstraction,
ratschan2007safety,segelken2007abstraction,sorea2004lazy}
and (b)~abstracting hybrid systems into hybrid systems with simpler dynamics~\cite{HARE,
dierks2007automatic,jha2007reachability,prabhakar2015hybrid,10.1007/978-3-319-65765-3_7}.
These two lines of work, although supported by  various automated tools  (e.g.,~\cite{ratschan2007safety,Henzinger1997,frehse2008phaver,frehse2011spaceex,chen2013flow}),
are difficult to be used in practice due to implicit and restrictive assumptions that they make on the structure of the hybrid systems under analysis.
Further, due to their limited scalability, they are inadequate for testing  CI-CPS models. 
For example,  Ratschan~\cite{ratschan2007safety} proposes an approach that took more than 10h to verify the RHB benchmark (a non-CI-CPS model also used in this paper). In contrast, our technique tests models instead of exhaustively verifying them.  Being black-box, our approach is agnostic to the modeling language used for MUT, and hence, is applicable to Simulink models irrespective of their internal complexities.  Further, as shown in our evaluation, our approach can effectively and efficiently test industrial CI-CPS models. 

There has been earlier work to combine CEGAR with testing  instead of model checking (e.g.,~\cite{ball2005abstraction,935473,Zutshi:2014:MSC:2656045.2656061,Zutshi:2015:FSP:2728606.2728648,Kroening2010,10.1007/978-3-540-70545-1_38,Kroening2010,10.1007/978-3-319-24953-7_35,7741019,10.1007/978-3-319-24953-7_35}). However, 
 based on  a recent survey on the topic~\cite{7741019},  \ourtool\ is the first approach that combines the ideas behind CEGAR with the system identification framework to develop an effective and efficient testing  framework for  CI-CPS models. Non-CEGAR based model testing approaches for CPS have been presented in the literature~\cite{dreossi2015efficient,Nghiem:2010:MTF:1755952.1755983,7963007,yaghoubi2017local,dreossi2015efficient,Arrieta2017,plaku2007hybrid,sankaranarayanan2012falsification,Bartocci2018} and are supported  by tools~\cite{7741019,ernst2019arch,staliro,donze2010breach,akazaki2018falsification,zhang2018two}. Among these,  we considered S-Taliro as a baseline for the reasons reported in Section~\ref{sec:ourtool}.

Zhang et al.~\cite{zhang2018two} reduce the number of simulations of the MUT by iteratively evaluating different inputs 
for short simulation times and by generating at each iteration the next input based on the  final state of the simulation. 
This approach assumes that the inputs are piecewise constants
and does not  support complex input profiles such as those used in our evaluation for testing our industry CI-CPS model. To reduce the simulation time of CI-CPS models, we  can manually simplify the models while preserving the behaviour needed to 
test the requirements of interest~\cite{alur:15,popinchalk2012improving}.
However, such manual simplifications are error-prone and reduce maintainability~\cite{arrieta2019pareto}. Further, finding an optimal balance between accuracy and execution time  is a complex task~\cite{Siddesh:2015:CSC:2898951}.

\vspace{-0.2cm}
\section{Conclusions}
\label{sec:conclusion}
We presented \ourtool,  a technique that combines  testing with an approximation-refinement loop to detect requirements violations in CI-CPS models.
We implemented \ourtool\ as a Matlab/Simulink application and compared its effectiveness and efficiency with the one of S-Taliro,  a state-of-the-art testing framework for Simulink models.
ARIsTEO finds $23.9\%$ more violations  than S-Taliro and finds those violations in $31.3\%$ less time than S-Taliro.
We evaluated 
the practical usefulness of ARIsTEO  on two versions of an industrial CI-CPS model to check three different requirements. 
ARIsTEO successfully triggered requirements violations in every case and required four hours on average for each violation, while S-Taliro failed to find any violations within four-hours. 
 
\begin{acks}
This work has received funding from the
\grantsponsor{erc}{European Research Council}{https://erc.europa.eu/} under the European Union's Horizon 2020
  research and innovation programme (grant agreement
 No~\grantnum{erc}{694277}), from QRA Corp, and from the University of Luxembourg (grant ``ReACP").\\
We thank our partners LuxSpace and QRA Corp for their support. 
\end{acks}

\balance

\end{document}